\documentclass[aps,prc,preprint,superscriptaddress,showpacs,amssymb]{revtex4}

\usepackage{graphicx}
\usepackage{amssymb}

\def\WARS{Institute of Experimental Physics, University of Warsaw, Ho\.za 69, 00-681 Warsaw, Poland}
\def\IFIC{Institut de F\'{\i}sica Corpuscular, Universitat de Valencia-CSIC, Dr Moliner 50, 46100 Burjassot, Spain}
\def\SUBA{SUBATECH, (CNRS/IN2P3, Ecole des Mines de Nantes, Universit\'e© de Nantes), Nantes, France}
\def\GANI{GANIL, IN2P3-CNRS, DSM-CEA, 14076 Caen Cedex 5, France}
\def\Barc{Grup de F\'{\i}sica de les Radiacions, Universitat Aut\`onoma de Barcelona 08193, Catalonia}
\def\Rez{Nuclear Physics Institute of the ASCR, 25068 \v{R}e\v{z}, Czech Republic}
\def\KVI{Kernfysisch Versneller Instituut, 9747 AA Groningen, The Netherlands}

\begin{document}

%\preprint{}

\title{Emission patterns of neutral pions in 40A MeV Ta+Au reactions}

\author{K.~Piasecki} \email[]{Krzysztof.Piasecki@fuw.edu.pl} \affiliation{\WARS}
\author{T.~Matulewicz} \affiliation{\WARS}
\author{N.~Yahlali} \affiliation{\IFIC}
\author{H.~Delagrange} \affiliation{\SUBA} \affiliation{\GANI}
\author{J.~D\'{\i}az} \affiliation{\IFIC}
\author{D.G.~d'Enterria} \thanks{Present address: CERN, CH-1211 Gen\`eve, Switzerland} \affiliation{\SUBA}
\author{F.~Fern\'{a}ndez} \affiliation{\Barc}
\author{A.~Kugler} \affiliation{\Rez}
\author{H.~L\"{o}hner} \affiliation{\KVI}
\author{G.~Mart\'{\i}nez-Garc\'{\i}a} \affiliation{\SUBA} \affiliation{\GANI}
\author{R.W.~Ostendorf} \affiliation{\KVI}
\author{Y.~Schutz} \thanks{Present address: CERN, CH-1211 Gen\`eve, Switzerland} \affiliation{\SUBA}\affiliation{\GANI}
\author{P.~Tlust\'y} \affiliation{\Rez}
\author{R.~Turrisi} \thanks{Present address: INFN-Padova, Via Marzolo 8, 35131 Padova, Italy} \affiliation{\GANI}
\author{V.~Wagner} \affiliation{\Rez}
\author{H.W.~Wilschut} \affiliation{\KVI}


\date{\today}

\begin{abstract}
Differential cross sections of neutral pions emitted in $^{181}$Ta+$^{197}$Au collisions
at a beam energy of 39.5A MeV have been measured with the photon spectrometer
TAPS. The kinetic energy and transverse momentum spectra of neutral pions cannot be
properly described in the framework of the thermal model, nor when the reabsorption of
pions is accounted for in a phenomenological model. However, high energy and high momentum
tails of the pion spectra can be well fitted through thermal distributions with unexpectedly
soft temperature parameters below 10 MeV.
\end{abstract}


\maketitle

\section{\label{intro}Introduction}

Neutral pion emission is a very sensitive probe of the processes taking place
in the interaction zone of two nuclei colliding at beam energies below 100A MeV.
This energy is well below the free nucleon-nucleon (NN) $\pi^0$ threshold of 280 MeV.
In consequence, the energy of produced mesons is limited, even with the help of
cooperative effect of Fermi motion. This energy limitation allows for the
experimental study of the energy spectrum and angular distribution of $\pi^0$ mesons 
to a large extent. In addition, the dominant two-photon decay channel
of this short lived meson makes possible the detection of neutral pions also at rest.
The number of detected neutral pions, superior to the previous measurement
\cite{Schu97,Marq94}, allows us to make more definitive conclusions about their
spectral properties.

With decreasing beam energy per nucleon, the particle production process
increasingly relies on the internal motion of nucleons and their correlations
(cooperative effects) as well as on the dynamical evolution of the colliding
system. The particles produced in these unfavourable conditions carry important
information about the collision process, as they remove a significant fraction
of the available energy.
Various models have attempted to describe these processes. Fermi motion, which
plays a crucial role in subthreshold particle production, is, of course,
incorporated in modern transport model calculations \mbox{({\it e.g.} \cite{Cas90})}.
These models include baryonic resonances, which serve as a temporary energy storage
enabling the production of mesons.
However, while the gross properties of subthreshold particle production
are relatively well explained by these models, the description of spectral
properties of neutral pions does not always provide satisfactory results.
While for Ar+Al collisions at 95A MeV a rather consistent description of $\pi^0$
emission was found \cite{Bada93}, the yield of neutral pions from Kr+Ni
collisions at 60A MeV was strongly underpredicted in the high energy part of
the spectrum \cite{Gudi96}. Phenomenological models of first-chance nucleon-nucleon
collisions have also been used to describe the energy scaling of the cross sections
\mbox{({\it e.g.} \cite{Meta93})}. In another extreme, purely thermal models
of particle emission have been applied to describe the interaction zone of two
colliding nuclei \cite{Zubk92}.
Strong reabsorption processes of pions in the nuclear medium have been
incorporated in both the transport and thermal models \cite{Bada93,Zubk92,BNV90}.

The production of neutral pions was studied with the TAPS spectrometer \cite{TAPS}
for a nearly symmetric heavy system $^{181}$Ta+$^{197}$Au at a beam energy of 
39.5A MeV, i.e. at 1/7 of the free NN $\pi^0$ production threshold energy. 
Particular attention was paid to removing all effects that could obscure the 
experimental data for such rarely produced particles. The contribution from cosmic 
radiation was reduced to below 1\% of the final neutral pion sample.

\section{\label{sec2}Experimental setup}

In the experiment performed at GANIL $^{181}$Ta$^{57+}$ ions were
accelerated to 39.5A MeV beam energy and delivered in bunches with a rate of
9.055 MHz. The nominal beam intensity was 10~enA, ca. 160 ions per pulse.
The Ta ions impinged on a \mbox{19.3~mg/cm$^2$} thick $^{197}$Au target.
The interaction probability per average beam pulse was about 7\%.

Photon pairs originating from $\pi^0$ decays were detected by the TAPS
electromagnetic calorimeter \cite{TAPS}, arranged in 6 blocks, each composed of
64 hexagonal modules in an 8$\times$8 honeycomb structure. The configuration of
the experimental setup was shown in Fig.~1 of Ref.~\cite{Ort05} (in our
measurement the inner detectors Dwarf Ball and SSD were removed). The blocks
covered a wide range of polar angles (between 45$^\circ$ and 170$^\circ$),
making up ca.~20\% of the full solid angle.
Each module consisted of a 25~cm-long BaF$_2$ crystal (corresponding to
12 radiation lengths) of hexagonal cross section with an inscribed radius of 2.95 cm.
A charged particle veto detector (CPV) made of plastic scintillator NE102A was mounted
in front of each module. Events with neutral pion candidates were selected by
a trigger, requiring activation of at least two modules by neutral particles,
in different TAPS blocks. An activation was defined by deposing energy of about 12 MeV,
a value not affecting the measurement of energy distribution of photons originating
from neutral pion decays.

\section{\label{sec3}Data analysis}

\begin{figure}
\includegraphics[width=8.6cm]{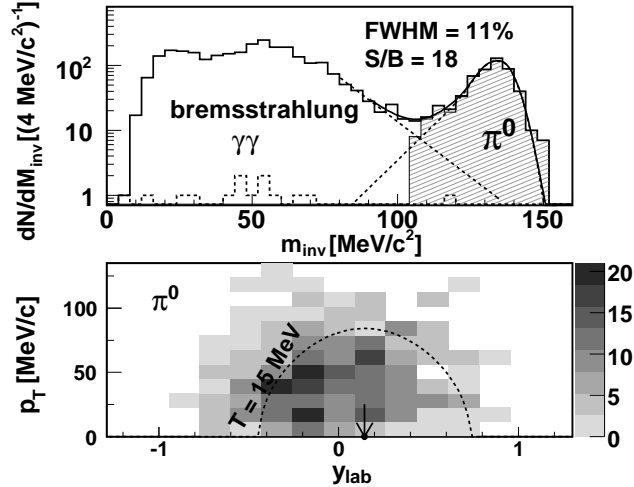}
\caption{\label{f-minv}Top: invariant mass distribution of photon pairs shown with 
  fit of an asymmetric Gaussian function to the $\pi^0$ peak and an exponential
  background. Hatched histogram: the $\pi^0$ candidates after the kinematical fit
  selection; dashed histogram: contamination from cosmic radiation.
  Bottom: Phase space distribution of the reconstructed $\pi^0$ in the $p_T$
  vs $y_{lab}$ plane; dashed curve for the average momentum of the $\pi^0$ obtained 
  from a Boltzmann distribution characterized by T=15~MeV; the arrow indicates the 
  $y_{CM}^{AA}$ (see text).}
\end{figure}

A single TAPS module measures the time, energy and electric charge state
(neutral or charged) of the incoming particle. The scintillation light response
of BaF$_2$ has two components, whose relative intensity differs for electrons
(and photons) compared to heavily-ionizing particles like protons and deuterons
\cite{Novo98}.
This feature, combined with the Time of Flight (ToF) information and the
charge veto signal, delivers nearly unambiguous identification of a photon
\cite{Marq95}.
This particle, traversing the BaF$_2$ module, induces an electromagnetic cascade
which usually spreads into adjacent modules, creating a cluster of active modules.
The properties of a particle were reconstructed from a cluster using
the ROSEBUD \cite{ROSE} analysis package. If a particle hits the border
of a TAPS block, part of the induced cascade can escape the detector.
Such a cluster was rejected by requiring that the maximum energy
deposition can not be in a border module (c.f. \cite{Marq95}).
Cosmic radiation was filtered out by a series of dedicated algorithms
\cite{Pias05,Matu93,Mart97}. They make use of differences between photon
and cosmic-ray induced clusters in the topology-based signatures:
the multiplicity of active modules in a cluster, the linearity of its
shape, and the intrinsic dispersion of the energy distribution in modules
calculated with respect to the most energetic one.
The invariant mass distribution of $\gamma\gamma$ pairs obtained is shown in the 
upper panel of Fig.~\ref{f-minv} and exhibits a prominent $\pi^0$ peak. The invariant
mass resolution (FWHM) of this peak is 11\%, in agreement with Monte-Carlo
simulations \cite{Marq95}. The observed shoulder for masses below $m_{\pi^0}$
is due to a partial escape of the electromagnetic cascade from a TAPS block
\cite{Matu90}. The high signal to background ratio S/B=18
allows selection of the $\pi^0$ candidates on an event-by-event basis.

A kinematical fit \cite{Korz00} was applied to each event to improve the
quality of the reconstruction of neutral pion momentum from the momenta of the
photons in a pair associated with the meson decay, taking advantage of the
known neutral pion mass. The parameters of this method, the invariant mass
threshold m$_{\gamma\gamma}^{min}$ and the maximum chi-square of the event
reconstruction $\chi^2_{max}$, were adjusted such that the distributions
of pulls of the three momentum components were in the best agreement with the
Normal distribution. The best fit parameters were found to be
\mbox{$m_{\gamma\gamma}^{min} \! = \! 105 ~MeV/c^2$} and \mbox{$\chi^2_{max} \! = 20$}.
The influence of their variation on the results presented in this paper
was added to the corresponding systematic errors. The bremsstrahlung
photon pairs from the Ta+Au collision zone were automatically rejected by
these conditions (mainly by the m$_{\gamma\gamma}^{min}$ cut). A total of
528 photon pairs satisfied all the above-mentioned filters and were assigned
to $\pi^0$ decays. The contribution of cosmic radiation was found to be on the level
of below one pair.

\begin{figure}
\includegraphics[width=8.6cm]{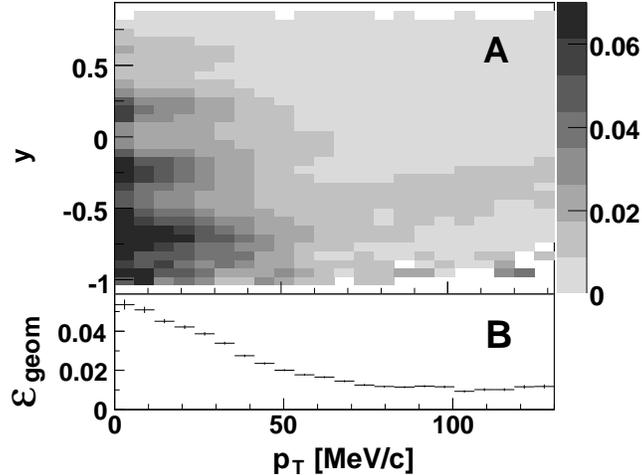}
\caption{\label{f-effpty}Top: efficiency of $\pi^0$ mesons as a function of 
rapidity and transverse momentum, obtained with the KANE code \cite{KANE}.
Bottom: estimation of geometric component of efficiency, simulated in frame of 
the PLUTO code \cite{PLUTO}. See text for details.}
\end{figure}

The phase space population of measured neutral pions is shown in the bottom
panel of Fig.~\ref{f-minv} in the plane of rapidity and transverse momentum,
where the midrapidity in the AA frame ($y_{CM}^{AA}$) is indicated by an arrow.
The dashed curve shows the average momentum of $\pi^0$ emitted from a source
obeying the Boltzmann distribution (Eq.~\ref{e-boltz0}) with a temperature of
15~MeV. This value is typical for the inverse slopes of momentum distributions
of neutral pions at beam energies around 40A MeV \cite{Schu97,Maye93,Pias02}.
The expected phase space distribution of neutral pions is well covered by 
the experimental data. To account for the detection efficiency, neutral pions 
were sampled from a Boltzmann distribution without collective radial flow of 
a particle 
\cite{Siem79}
\begin{equation}
  \frac{d^2\sigma}{dp_T dy} \sim p_T \cdot E \cdot \exp(-E / T) ,
\label{e-boltz0}
\end{equation}
where $T$ is the temperature of the source, $E = m_T c^2 \cdot cosh (y)$ is used and 
$m_T$ is the transverse mass $m_T c^2 = \sqrt{(p_T c)^2+(m_T c^2)^2}$ \cite{PDG}. 
The response of the TAPS 
apparatus was simulated using the GEANT3-based \cite{GEANT} KANE code \cite{KANE}.
The source temperatures were varied between 6 and \mbox{24 MeV}, i.e. a range of 
inverse slopes of transverse momentum distributions found to be characteristic
for beam energies between 25A and 95A MeV \cite{Schu97,Maye93,Pias02,Tym05,Yahl05,Tym06}.
The efficiency $\varepsilon$ of neutral pions was investigated as a function of 
their rapidity and transverse momentum (see the upper panel of Fig.~\ref{f-effpty}).
In the region of $p_T \lesssim $~60~MeV/c the shape of $\varepsilon$ changes rapidly.
Oscillations in rapidity stem from the geometric configuration of the TAPS blocks.
In order to verify that the sharp drop of efficiency at lower transverse 
momenta is also due to the geometric setup of TAPS, a thermal source of $\pi^0$ 
mesons was generated within the PLUTO \cite{PLUTO} code, and boosted according to
the NN frame in the colliding system analysed. An efficiency $\varepsilon_{geom}$
was constructed by selecting events, where both photons from a $\pi^0$ decay were 
emitted towards the TAPS blocks. The profile of $\varepsilon_{geom}$ 
as a function of $p_T$, shown in the lower panel of Fig.~\ref{f-effpty}, demonstrates 
that the TAPS geometry is the driving cause of the drop observed in the upper panel of
this figure. It also provides a cross-check of the GEANT-based analysis.
In the regions of phase space around the limits of observed data some dependency 
of the efficiency on $T$ was found. The existence of this dependency in conjunction with
the profile of measured $p_T$ spectrum (c.f. Fig.~\ref{f-pt}) is the reason behind 
using as an input to the KANE efficiency calculation a thermal distribution with a 
broad range of values of $T$ parameter, rather than a homogenous spectrum in $p_T-y$ 
phase space. Variations of parameters extracted in the present analysis due to the 
uncertainty of the efficiency map were added to the relevant systematic errors. 
As the efficiency changes quite sharply in some regions of $p_T$ and $y$,
in order to obtain the weight for every reconstructed neutral pion, $\varepsilon$ 
was linearly interpolated between neighbouring bins.

The overall $\pi^0$ acceptance of the apparatus and the applied analysis methods 
was found to be 0.36\%.

\section{\label{sec4}Results}

In the previous measurement of Ta+Au collisions at 39.5A MeV, performed
by the TAPS collaboration, about 100 $\pi^0$ mesons were reconstructed
\cite{Schu97,Marq94}. The neutral pion cross section was found to be
\mbox{$\sigma_{\pi^0}$~=~2.2$\pm$0.3~$\mu$b} and the transverse momentum spectrum
was described by a Boltzmann function characterized by a temperature of
16$\pm$4 MeV. The cross section for bremsstrahlung photons of energies above 30~MeV,
\mbox{$\sigma_\gamma$~=~6.9$\pm$0.7~mb}, was also obtained, thus the ratio of
$\sigma_{\pi^0} / \sigma_\gamma$ was found to be (3.2$\pm$0.5)$\cdot 10^{-4}$.

The analysis presented here is based on a number of identified pions 5
times larger compared to the previous measurement. Therefore, we are now able
to study the properties of produced pions in more detail. As a cross-check
with the previous measurement, the $\sigma_{\pi^0} / \sigma_\gamma$ ratio was
obtained according to the following formula:
\begin{equation}
  \frac{\sigma_{\pi^0}}{\sigma_\gamma} =
    \frac{\sum_{i=1}^{N_\pi}1/ \varepsilon^\pi_i}
         {\sum_{i=1}^{N_{\gamma}}1/ \varepsilon^\gamma_i} ,
\end{equation}
where $\varepsilon^{\pi/\gamma}$ are the detection efficiencies of neutral pions
and photons, and the summation runs over $N_{\pi/\gamma}$ accepted particles.
The relevant efficiency maps were simulated using the KANE code.
The ratio obtained is found to be (4.5$\pm$0.4)$\cdot 10^{-4}$, in 2$\sigma$
agreement with the value found in the previous measurement. For an absolute
normalization of the neutral pion distributions presented here, the above mentioned
cross section of 2.2$\pm$0.3~$\mu b$ from Refs.~\cite{Schu97,Marq94} was used.

\begin{figure}
\includegraphics[width=8.6cm]{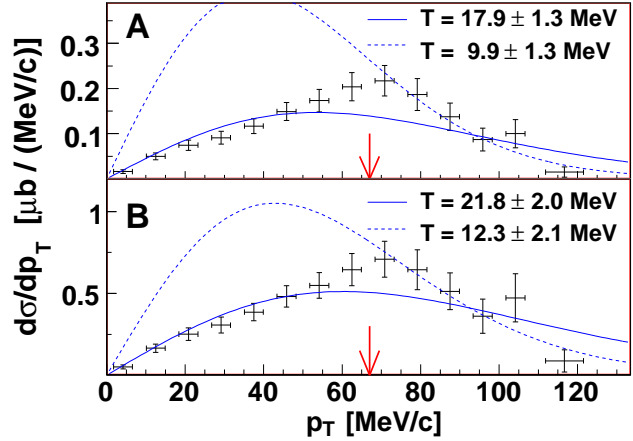}
\caption{\label{f-pt}(Color online) Transverse momentum distribution of neutral pions emitted from
  Ta+Au collisions at a beam energy of 39.5A~MeV, for a $\Delta y = \pm$0.45 window
  around $y_{AA}^{CM}$, without (a) and including (b) the model of absorption in 
  nuclear matter (see text). Thermal fits to the whole spectrum (solid lines) 
  and to the high $p_T$ range (dotted lines) are shown. Arrows indicate the lower
  bound of the fit to a high $p_T$ tail of the spectrum.}
\end{figure}
For a narrow rapidity range $\Delta y$ around $y_{CM}^{AA}$, the Boltzmann
distribution expressed by Eq.~\ref{e-boltz0} can be approximated by:
\begin{equation}
 \frac{d\sigma}{dp_T} \sim p_T \cdot m_T \cdot \exp
 \left(-\frac{m_T c^2 \left< cosh(y) \right>_{\Delta y}}{T} \right) ,
\label{e-boltz2}
\end{equation}
To fit this function to the 
experimental data pions from the $\Delta y$~=~$\pm$0.45 range were selected. This 
choice of rapidity range is a balance between the requirement of reasonable statistics
and the reliability of approximation assumed in Eq.~\ref{e-boltz2}. Variations
of the results presented below due to a choice of $\Delta y$ were added to the
systematic errors. The fit gives an inverse slope of 17.9$\pm$1.3$^{+0.7}_{-1.3}$ MeV
(see solid curve in Fig.~\ref{f-pt}(a) and Table~\ref{tab-sum}). However, 
a $\chi^2 / \nu$ is found to be 2.5, and the auto-correlations in the residuals 
of neighbouring bins can be observed.
The same discrepancy is also observed for the kinetic energy distibution, shown in
Fig.~\ref{f-ek} and discussed below. The fit of Eq.~\ref{e-boltz2}
to the high $p_T$ part ($p_T \! > 67 MeV/c$) delivers an inverse slope parameter of
\mbox{$T=$~9.9$\pm$1.4$^{+1.0}_{-0.8}$~MeV}, a value considerably lower than the 
one resulting from the fit in a full $p_T$ range.
However, in this case the low momentum part of the experimental spectrum 
is strongly overpredicted by the extrapolation of the fit. 
This depletion correlates with the decline of efficiency with $p_T$, as
shown in the upper panel of Fig.~\ref{f-pt}. However, as reported in section~\ref{sec3},
an additional simulation verified that the positioning of TAPS blocks was solely
responsible for the mentioned drop of efficiency.
The findings reported above strongly suggest that the distribution of transverse 
momentum of neutral pions cannot be fitted globally with the Boltzmann's thermal
function.
As shown in Table~\ref{tab-ptfits}, this trend has been observed in a range 
of systems investigated by the TAPS collaboration at beam energies below 
100A~MeV~\cite{Schu97,Maye93,Pias02,Tym05,Yahl05,Tym06}. 
The $p_T$ spectra of neutral pions from other experiments were taken as they
were published and fitted both to the full range and with a common lower bound of 60~MeV/c
with the function 
\begin{equation}
f(p_T) \sim p_T \cdot m_T \cdot exp(- m_T /T)
\label{e-boltzeff}
\end{equation}
In spite of somewhat different ranges of rapidities accessible to those experiments,
in all reported cases the slope obtained globally was larger than the one fitted 
to the high $p_T$ tail of distribution. 
Therefore, as for the current data, extrapolations of Boltzmann functions fitted 
to the high $p_T$ tails overpredicted the measured pion yield at lower transverse momenta.
In the measurements performed in the eighties at 25A and 35A~MeV \cite{Stac86,Youn86},
no depletion of yield was found. However, these experiments 
were performed with lead-glass spectrometers, which allowed for an invariant mass 
resolution of only 40\%. Also, the experimental setup used in these measurements 
did not allow for the application of advanced procedures for the rejection of 
cosmic-ray background, employed in the analysis presented in our paper.

\begin{table}
\caption{Systematics of inverse slopes obtained by fitting the Eq.~\ref{e-boltzeff}
to the transverse momentum distributions of neutral pions at beam energies below 100A~MeV
from recent experiments. Global fits ($p_T$~$>$~0~MeV/c) and high $p_T$ ($>$~60 MeV/c)
are shown. In the cases where authors used different functions, or lower fit bounds 
for fitting the tail was different than 60~MeV/c, the data were extracted and refitted.}
\begin{ruledtabular}
\label{tab-ptfits}
\begin{tabular}{|c|c|c|c|c|}
System  & T$_{beam}$~[A~MeV] & T~($p_T$~$>$~0~MeV/c) [MeV] & T~($p_T$~$>$~60 MeV/c) [MeV] &      Reference     \\
\hline
 Ar+Au  &      25          &       13.3$\pm$2.5            &          6.4$\pm$1.9         & \cite{Yahl05}, prelim. \\
 Ar+Au  &      35          &       13.0$\pm$0.9            &         11.6$\pm$0.9         & \cite{Yahl05}, prelim. \\
 Ta+Au  &      39.5        & 17.4$\pm$1.3$^{+0.7}_{-1.3}$  & 11.0$\pm$1.2$^{+1.0}_{-0.8}$ &     this work      \\
 Xe+Au  &      44          &       14.6$\pm$1.0            &         12.7$\pm$1.0         &    \cite{Maye93}   \\
 Kr+Ni  &      60          &       17.0$\pm$0.5            &         15.8$\pm$0.2         &    \cite{Schu97}   \\
 Ar+C   &      60          &       16.2$\pm$0.7            &         13.3$\pm$1.0         & \cite{Tym05,Tym10} \\
 Ar+Ni  &      60          &       16.6$\pm$0.6            &         11.9$\pm$0.3         & \cite{Tym05,Tym10} \\
 Ar+Ag  &      60          &       18.3$\pm$2.8            &         16.3$\pm$2.7         & \cite{Tym05,Tym10} \\
 Ar+Au  &      60          &       17.2$\pm$0.8            &         12.7$\pm$0.7         & \cite{Tym05,Tym10} \\
 Ar+C   &      95          &       16.1$\pm$0.2            &         14.2$\pm$0.1         & \cite{Tym05,Tym10} \\
 Ar+Al  &      95          &       18.4$\pm$0.1            &         16.4$\pm$0.1         & \cite{Tym05,Tym10} \\
 Ar+Ag  &      95          &       16.8$\pm$0.2            &         14.1$\pm$0.2         & \cite{Tym05,Tym10} \\
 Ar+Au  &      95          &       15.2$\pm$0.2            &         13.6$\pm$0.2         & \cite{Tym05,Tym10} \\
\end{tabular}
\end{ruledtabular}
\end{table}

\begin{figure}
\includegraphics[width=8.6cm]{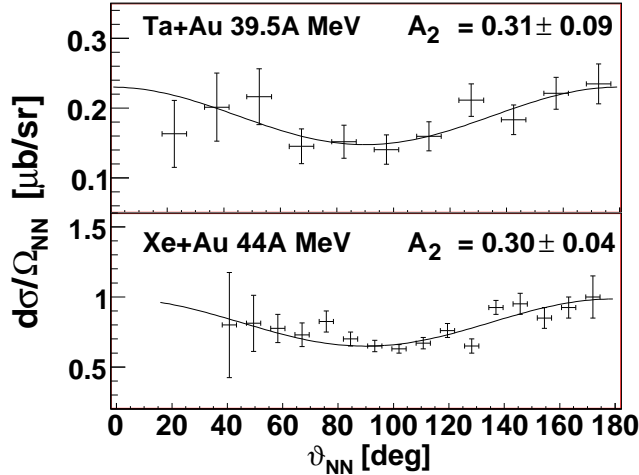}
\caption{\label{f-thet}Experimental angular distribution of neutral pions emitted from
  Ta+Au at 39.5A MeV shown in the NN frame, compared to Xe+Au at 44A MeV \cite{Maye93}.
  Curves represent the best fits of the second order Legendre polynomial (see text).}
\end{figure}

To estimate the effect of the shadowing of pions by nuclear matter, a simple geometrical
model \cite{Holz96,Tym02} was applied. It assumes $\pi^0$ production from
a random point inside a static collision zone defined by the maximum overlap
of two spheres at a random impact parameter. Mesons are produced in a
thermalized source of temperature $T_{model} = 12$~MeV and emitted with
an angular distribution:

\begin{equation}
f(\vartheta) = 1 + A_2 \cdot P_2(cos\vartheta) ,
\label{e-thet}
\end{equation}
where $P_2$ is the second order Legendre polynomial and \mbox{$A_2 = 0.05$} is the
anisotropy parameter\footnote{50\% variations of $T_{model}$ have negligible
influence on the obtained corrections. Variations due to changes of $A_2$ within
a \mbox{(-0.5,}\mbox{0.5)} interval are small and were included in the
systematic errors.}.
This model uses the momentum-dependent mean free path of a neutral pion \cite{Meh84}.
However, it must be noted that neither pion scattering nor the dynamics of the 
collision is included in this model.

The primordial $p_T$ distribution estimated by this model is shown in
\mbox{Fig.~\ref{f-pt}(b).} A fit of Eq.~\ref{e-boltz2} performed in the whole
range of transverse momenta delivers the inverse slope of 21.8$\pm$2.0~MeV and a
$\chi^2 / \nu$~=~1.6. 
%Compared to the $\chi^2 / \nu$ value of 2.5 obtained 
%without accounting for the pion absorption, an application of this model brings 
%the distribution somewhat closer to the thermal shape.
This is an improvement compared to $\chi^2 / \nu$~=~2.5 for the measured data, 
so an application of this model brings the distribution somewhat closer to the 
thermal shape. 
However, auto-correlations in the residuals of neighbouring bins still persist.
Fitting the high $p_T$ tail with a Boltzmann distribution gives a considerably lower
inverse slope of \mbox{$T=$~12.3$\pm$2.1$^{+1.8}_{-1.1}$~MeV}. However, the fitted
function still strongly overestimates the measured yield at lower $p_T$ values.
In the scenario of thermal equilibrium established during the collision, and the
system cooling gradually, neutral pions of high $p_T$ or kinetic energies should 
be rather characterized by highest temperatures. However, this approach is in
contradiction with the findings presented above.
The mismatches demonstrated above suggest that neutral pions produced and emitted 
from the collision zone may not be thermally equilibrated, even if the absorption
of neutral pions is accounted for in frame of the above mentioned model.

\begin{turnpage}
\begin{table*}
\caption{Summary of inverse slopes and polar anisotropy parameters. Also, shown
are the estimates of the primordial values of these parameters, calculated within the
$\pi^0$ absorption model described in the text. For the contributions to the systematic
errors, see text.
The global fits ($p_T>0$, $E_{kin}>0$) could not describe the spectra well.}
\begin{ruledtabular}
\label{tab-sum}
\begin{tabular}{|c|c|c||c|c|}
\multicolumn{1}{|c|}{Observable}                        &
\multicolumn{2}{c||}{result from measured spectrum}     &
\multicolumn{2}{c|}{result from primordial spectrum}  \\
\hline
\multicolumn{1}{|c|}{T from $p_T$ for $|y-y^{CM}_{AA}|\! <$~0.45} &
\multicolumn{1}{c|}{for $p_T \! > $~0~MeV/c}            &
\multicolumn{1}{c||}{for $p_T \! > $~67~MeV/c}          &
\multicolumn{1}{c|}{for $p_T \! > $~0~MeV/c}            &
\multicolumn{1}{c|}{for $p_T \! > $~67~MeV/c}            \\
\cline{2-5}
\multicolumn{1}{|c|}{}                                  &
\multicolumn{1}{c|}{17.9$\pm$1.3$^{+0.7}_{-1.3}$ MeV}   &
\multicolumn{1}{c||}{9.9$\pm$1.4$^{+1.0}_{-0.8}$ MeV}   &
\multicolumn{1}{c|}{21.8$\pm$2.0 MeV}                   &
\multicolumn{1}{c|}{12.3$\pm$2.1$^{+1.8}_{-1.1}$ MeV}    \\
\hline
\multicolumn{1}{|c|}{T from $E_{kin}$ for $E_{kin}\! >$~37~MeV}&
\multicolumn{2}{c||}{8.7$\pm$1.6$^{+0.1}_{-1.3}$ MeV}   &
\multicolumn{2}{c|}{10.8$\pm$2.2$^{+0.2}_{-1.7}$ MeV}   \\
\hline
\multicolumn{1}{|c|}{$A_2$}                             &
\multicolumn{2}{c||}{0.31$\pm$0.09$^{+0.05}_{-0.02}$}   &
\multicolumn{2}{c|}{0.09$\pm$0.09$^{+0.03}_{-0.02}$}    \\
\end{tabular}
\end{ruledtabular}
\end{table*}
\end{turnpage}

The experimental polar angle distribution of neutral pions in the NN frame,
(see Fig.~\ref{f-thet}) exhibits an anisotropy with less preference for emission
in sideward directions. The lack of strong asymmetry between
forward and backward directions can be explained by the near-symmetry
of the colliding system. The pronounced sideward absorption can be qualitatively
explained by the longer path of a pion in the colliding nuclear matter due to
mutual shadowing by the heavy collision partners.
The fit of Eq.~\ref{e-thet} to the experimental data yields an anisotropy
parameter $A_2$~=~0.31$\pm$0.09$^{+0.05}_{-0.02}$.
In an attempt to reconstruct the primordial polar angle distribution of neutral pions,
the above-mentioned $\pi^0$ absorption model was used. To parametrize the primordial
spectrum, a dipolar distribution of the form of Eq.~\ref{e-thet} was assumed,
with an anisotropy parameter $A_2^{prim}$. The angular spectrum of pions which
traversed the hadronic matter was compared to the spectrum observed
by TAPS. The best fit to the experimental data was obtained for
$A_2^{prim}$~=~0.09$\pm$0.09$^{+0.03}_{-0.02}$, which exhibits a near isotropic
emission pattern.
Despite the meager experimental knowledge of $A_2^{prim}$ values for collisions
of symmetric systems at comparable beam energies, an average anisotropy
parameter $A_2^{prim} =$~0.32$\pm$0.05 for the primordial distribution of all
hitherto investigated systems was found \cite{Tym05,Tym02}. This value is quite
higher than the one obtained for Ta+Au. However, several values are also beyond
the range of the global average quoted above.
A fit of Eq.~\ref{e-thet} to the angular distribution of neutral pions emitted from
$^{129}$Xe+$^{197}$Au collisions at 44A~MeV \cite{Maye93} (see bottom panel of
\mbox{Fig. \ref{f-thet}}) provides an anisotropy parameter of $A_2$~=~0.30$\pm$0.04. An
estimation of this parameter for the primordial pions done in the framework of the
above mentioned $\pi^0$ absorption model gives $A_2^{prim}$~=~0.10$\pm$0.04. Both
values are in very good agreement with results for Ta+Au collisions obtained in
this work.

\begin{figure}
\includegraphics[width=8.6cm]{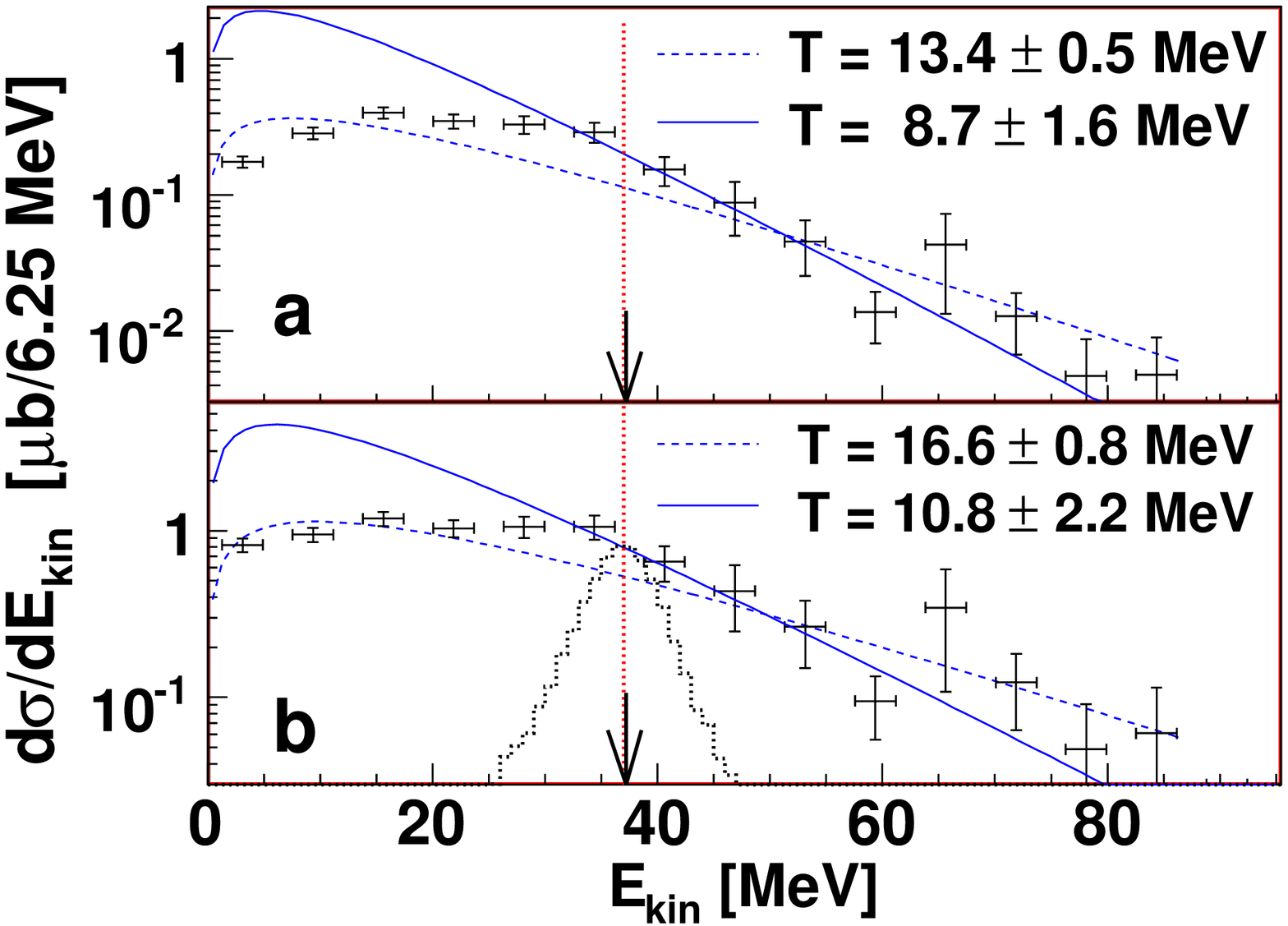}
\caption{\label{f-ek}(Color online) Kinetic energy spectrum of neutral pions, without (upper panel)
  and including (lower panel) the correction based on the $\pi^0$ absorption
  model (see text). Curves represent fits of Maxwellian distribution to the whole
  spectrum (solid lines) and to the high $E_{kin}$ part (dotted lines; vertical
  lines mark the lower fit bounds. Arrows indicate the kinematical limit for the
  $NN \rightarrow NN\pi^0$ reaction including Fermi motion. The TAPS
  response function for pions emitted with this limiting energy is shown by the
  dashed lines.}
\end{figure}

The kinetic energy distribution in the NN frame has also been investigated.
Global fitting by a Maxwellian distribution of the form 
$\sim pE \cdot exp(-E/T)$ [see Fig.~\ref{f-ek}(a)] 
results qualitatively in the same picture as in the case of $p_T$ spectrum: high value 
of $\chi^2 / \nu$~=~4.5 and auto-correlations in the residuals of neighbouring bins.
A fit of the same function to the high energy tail \mbox{($E_{kin} \! > \! 37$~MeV)}
gives an inverse slope of \mbox{T~=~8.7$\pm$1.6$^{+0.1}_{-1.3}$~MeV}. Similarly to
the case of the $p_T$ distribution, the application of the above mentioned $\pi^0$
absorption model results in a slightly increased inverse slope of the high energy 
tail of the estimated primordial spectrum, \mbox{T~=~10.8$\pm$2.2$^{+0.2}_{-1.7}$~MeV},
and the depletion of the yield in the low $E_{kin}$ part of the spectrum cannot 
be removed \mbox{[see Fig.~\ref{f-ek}(b)]}. 

The constructive superposition of the beam momentum and two Fermi momenta of
nucleons ($p_F$~=~270~MeV/c) from the colliding nuclei enables the
production of pions up to kinetic energy of $E_{max}^{NN}$~=~37~MeV. However,
the measured kinetic energy spectrum of neutral pions is spread far beyond this 
value. The experimental smearing of this upper energy limit, simulated using the
KANE code \cite{KANE}, and shown in Fig.~\ref{f-ek}, cannot explain
the extent of the high $E_{kin}$ shoulder.
One possible hypothesis leading to an explanation of this effect could be
the production of pions in channels different from the process $NN\rightarrow NN\pi^0$,
like the multi-step process
\mbox{i) $NN \rightarrow N\Delta, \Delta N \rightarrow NN \pi^0$}
or the cooperative one \mbox{ii) $NNN \rightarrow NNN \pi^0$}. To
estimate the kinetic energy limit for each of these processes, which would also
take into account the momentum conservation of the reaction products involved, the
collisions studied were simulated with the help of the PLUTO code \cite{PLUTO}.
A kinetic energy of 39.5A~MeV was assigned to a ''beam nucleon''. The Fermi
motion of nucleons was included. In the second step of channel i) all the
available energy was assumed to be transferred to the $\Delta$(1232) resonance
mass, an assumption justified by the rise of the $\Delta$ production
probability with a resonance mass in the region well below
$\left< m_\Delta \right>$. The $\Delta$ resonance interacted with another
nucleon of Fermi gas, originating randomly from the target or beam nuclei.
We found that the kinetic energy limits amount to around 75~MeV for both
investigated channels.
This may be a hint that the investigated channels could play a role in
explaining the observed presence of neutral mesons above the $E_{max}^{NN}$ limit.
On the other hand, calculations within framework of the Dubna Cascade
Model, assuming $\pi^0$ production in NN collisions, and employing
inelastic $\pi$N scattering \cite{Gudi96}, systematically underestimated
the measured yield in the high energy tail of the spectrum. More
systematic investigations of possible dynamical processes are needed to explain
the experimental findings.

\section{\label{sec5}Conclusions}

Neutral pions produced in the Ta+Au collision zone at 39.5A~MeV were
measured using the TAPS electromagnetic calorimeter. 
Their transverse momentum distribution selected around midrapidity and
kinetic energy spectrum deviate from a thermal shape.
These deviations cannot be removed by the application of a phenomenological 
model of $\pi^0$ absorption in nuclear matter.
The kinetic energy spectrum reaches far higher values than the limit for the
$NN \rightarrow NN \pi^0$ channel, including constructive boost of Fermi motion.
It is suggested that multi-step or cooperative processes could be possible 
sources of the neutral pion production.
Nonetheless, the tails of the $p_T$ and $E_{kin}$ spectra were in agreement 
with Boltzmann distributions, yielding inverse slope parameters of
\mbox{$T=$~9.9$\pm$1.4$^{+1.0}_{-0.8}$~MeV} and
\mbox{T~=~8.7$\pm$1.6$^{+0.1}_{-1.3}$~MeV}, respectively. 
%These values are
%significantly below those reported previously at similar beam energy range.
The polar angle distribution of the measured pions reveals a dipolar anisotropy
characterized by $A_2$=0.31$\pm$0.09$^{+0.05}_{-0.02}$. However, the estimated
asymmetry of the primordial angular distribution, calculated within the $\pi^0$
absorption model, turned out to be almost isotropic 
($A_2^{prim}$~=~0.09$\pm$0.09$^{+0.03}_{-0.02}$).

\begin{acknowledgments}

We thank the GANIL accelerator staff for delivering a high quality beam.
This research was supported in part by the Polish State Commitee for
Scientific Research (KBN) under grant \mbox{2P03B 102 25}, and by the
Spanish Ministerio de Educaci\'on y Ciencia under contract FPA2006-12120-C03-02.

\end{acknowledgments}

% If in two-column mode, this environment will change to single-column
% format so that long equations can be displayed. Use
%\begin{widetext}
%\end{widetext}

% tables should appear as floats within the text
% The ruledtabular enviroment adds doubled rules to table and sets a
% reasonable default table settings.
% Use the table* environment to get a full-width table in two-column
% Add \usepackage{longtable} and the longtable (or longtable*}
% environment for nicely formatted long tables. Or use the the [H]
% placement option to break a long table (with less control than 
% \begin{table}%[H] add [H] placement to break table across pages
% \begin{ruledtabular}
% Lines of table here ending with \\
% \end{tabular}
% \end{ruledtabular}
% \end{table}

% Surround table environment with turnpage environment for landscape
% table
% \begin{table}
% \begin{ruledtabular}
% \end{tabular}
% \end{ruledtabular}
% \end{table}


\begin{thebibliography}{00}
\bibitem{Tym10}  K.~Tymi\'nska, priv. comm. Preliminary results for $\pi^0$ measured at 60A MeV 
were presented in \cite{Pias02}
\end{thebibliography}
\end{document}